\documentclass[%
 reprint,
 superscriptaddress,
 amsmath,amssymb,
 aps,
prl
]{revtex4-2}

\usepackage{graphicx}
\usepackage{dcolumn}
\usepackage{bm}
\usepackage{hyperref}

\newcommand{\br}{\bm{r}}

\begin{document}


\title{Electrical Conductivity of Iron in Earth's Core from Microscopic Ohm's Law}


\author{Kushal Ramakrishna}
\email{k.ramakrishna@hzdr.de}
\affiliation{Center for Advanced Systems Understanding (CASUS), D-02826 G\"orlitz, Germany}
\affiliation{Helmholtz-Zentrum Dresden-Rossendorf (HZDR), D-01328 Dresden, Germany}

\author{Mani Lokamani}
\affiliation{Helmholtz-Zentrum Dresden-Rossendorf (HZDR), D-01328 Dresden, Germany} 

\author{Andrew Baczewski}
\affiliation{Center for Computing Research, Sandia National Laboratories, Albuquerque NM 87185 USA}

\author{Jan Vorberger}
\affiliation{Helmholtz-Zentrum Dresden-Rossendorf (HZDR), D-01328 Dresden, Germany}

\author{Attila Cangi} 
\email{a.cangi@hzdr.de}
\affiliation{Center for Advanced Systems Understanding (CASUS), D-02826 G\"orlitz, Germany}
\affiliation{Helmholtz-Zentrum Dresden-Rossendorf (HZDR), D-01328 Dresden, Germany}





\date{\today}

\begin{abstract}
Understanding the electronic transport properties of iron under high temperatures and pressures is essential for constraining geophysical processes. The difficulty of reliably measuring these properties under Earth-core conditions calls for sophisticated theoretical methods that can support diagnostics. We present results of the electrical conductivity within the pressure and temperature ranges found in Earth’s core from simulating microscopic Ohm’s law using time-dependent density functional theory. Our predictions provide a new perspective on resolving discrepancies in recent experiments.
\end{abstract} 

\maketitle


\textit{Introduction.~---~} 
Iron is the most abundant element by mass on planet Earth~\cite{doi:10.1021/cb300323q}. It makes up the majority of its liquid outer core and solid inner core~\cite{PhysRevB.50.6442,jeanloz1990nature}, which is exposed to temperatures of about 6000~K and pressures of about 300~GPa. Understanding the properties of iron under these extreme conditions is of great geophysical importance as they determine the internal structure of Earth. 
Likewise, the behavior of iron under elevated temperatures and pressures also plays a major role in materials science. A wide range of novel steel micro-structures can be produced with minor changes in composition and proper thermal treatment of iron-based alloys~\cite{pepperhoff2013constitution, morard2014properties}.  

While the iron phase diagram ~\cite{PhysRevLett.50.130,anderson1986properties,pourovskii2019electronic,hwang2020subnanosecond,white2020time,kruglov2021crystal} and the equation of state ~\cite{garai2011pvt,andrews1973equation,dorogokupets2017thermodynamics,anderson1986properties,pourovskii2019electronic,zhang2010static,dewaele2006quasihydrostatic,sha2010first,zhang2010static,grant2021equation,PhysRevLett.124.165701} have been well studied in the past decades, the phase diagram near Earth's inner core conditions and melt line remain inconclusive~\cite{ross1990theory, matsui1997case,belonoshko2003stability,PhysRevB.53.14063,vovcadlo2000ab,PhysRevB.50.6442,PhysRevB.98.224301,belonoshko2017stabilization}. 

Beyond equation-of-state data, transport properties of iron, such as its electrical and thermal conductivity, are intricately related to the geophysical dynamics that take place in the planetary interior. 
Most prominently, the heat flux between the planetary core and mantle drives the dynamo action~\cite{labrosse2003thermal,stacey2007revised} which generates the magnetic field of the Earth. Furthermore, the cooling rate of the core-mantle boundary is used to estimate the age of our planet~\cite{gubbins2004gross}.  

However, information on electronic transport properties under the conditions of Earth's core is sparse. This is due to the difficulties of achieving accurate measurements in experiments.
These are commonly performed in diamond-anvil  cells (DAC)~\cite{gomi2013high,ohta2016experimental,konopkova2016direct}, with wire-heating techniques~\cite{PhysRevB.42.6485,beutl1994thermophysical}, and using static and dynamic shock-compression~\cite{keeler1969electrical,gathers1983thermophysical,gathers1986dynamic}.
Shock compression techniques combined with x-ray Thomson scattering (XRTS)
provide diagnostics for the dynamical and static conductivity, which have been used for warm dense dense conditions in simple metals~\cite{sperling2015free}.
Most recently, terahertz transmission measurements of the time-resolved electrical conductivity in warm dense gold~\cite{chen2021ultrafast} have shown promise as a viable approach for further probing transport properties under extreme conditions.  

Notably, the aforementioned experiments using laser-heated diamond-anvil cells~\cite{ohta2016experimental,konopkova2016direct} have led to a controversy in the measurement of the electronic transport properties in iron at Earth core conditions~\cite{dobson2016earth}. Ohta \textit{et al.}~\cite{ohta2016experimental} infer a thermal conductivity of 90~Wm$^{-1}$K$^{-1}$ by measuring the electrical resistance of iron wires and converting it into a thermal conductivity using the Wiedemann-Franz law. Kon{\^o}pkov{\'a} \textit{et al.}~\cite{konopkova2016direct} 
measured the thermal diffusion rate for heat transferred between the ends of solid iron samples, inferring a thermal conductivity of 30~Wm$^{-1}$K$^{-1}$ from agreement with a finite-element model. The discrepancy in these measurements have deep implications for predictions of the age of the Earth leading to vastly different results ranging from 700 million to 3 billion years old~\cite{dobson2016earth}. 
Since the uncertainty in the electrical conductivity, both from experiment and theory, is so high, reliable knowledge about the fundamental processes generating Earth's magnetic field is lacking as well.
Due to the disagreement among existing experimental data, computational modeling of electronic transport properties under extreme conditions is indispensable in supporting current and future research, in particular, to further probe the earth-core conditions~\cite{berrada2021review}.  

The pioneering theoretical works use the Kubo-Greenwood (KG) formula~\cite{kubo1957statistical} and have been applied in modeling degenerate plasma states~\cite{PhysRevE.66.025401,plagemann2012dynamic,PhysRevLett.118.225001} and liquid metals~\cite{pozzo2011electrical,de2012electrical,PhysRevB.85.184201}. These evaluate the KG formula using the Kohn-Sham (KS) orbitals, eigenvalues, and occupation numbers obtained from density functional theory (DFT) calculations at finite electronic temperature~\cite{PhysRev.136.B864,KoSh1965,Mermin_1965}.      
Most recently, Korell \textit{et al.}~\cite{korell2019paramagnetic} have investigated the effects of spin-polarization on the electrical conductivity obtained from the KG formula, specifically for the paramagnetic state of liquid iron. This formulation has also been used for evaluating the electrical and thermal conductivity of iron and iron-silicon mixtures at Earth core conditions~\cite{pozzo2012thermal,pozzo2014thermal}.  
The direct use of KS quantities in the KG formula, however, is based on the response function evaluated with the KS orbitals with no interaction kernel to capture collective effects. This is especially relevant under the conditions when the electrons in iron are strongly correlated~\cite{pourovskii2016fermi}.

Improvements upon transport properties of iron at extreme conditions computed using DFT are possible by means of dynamical mean field theory (DMFT)~\cite{RevModPhys.78.865}. This approach takes into account the on-site Coulomb interaction, which is particularly strong for the localized \textit{3d} electrons in iron~\cite{belozerov2014coulomb}. The net conductivity consists of electron-lattice scattering usually evaluated with the KG formula and the electron-electron scattering (EES) evaluated with DMFT which takes into account the electronic correlations and the thermal disorder.
EES contributions in HCP iron are reported to be insignificant compared to electron-lattice scattering at Earth core conditions, but have important contributions to the total thermal conductivity~\cite{pourovskii2017electron}.

A viable alternative to the KG formula is linear response time-dependent density functional theory (LR-TDDFT)~\cite{GrKo1985}. 
While KS orbitals are used to calculate a non-interacting response function, the Hartree and exchange-correlation (XC) kernels are used to obtain an interacting response function that includes electron-electron correlations. Furthermore, LR-TDDFT yields full wavenumber and frequency resolved transport properties.
This method has recently been assessed in detail for solid and liquid aluminum~\cite{ramakrishna2020firstprinciples,dornheim2020effective}. However, the calculations in LR-TDDFT~\cite{doi:10.1146/annurev.physchem.55.091602.094449} rely on matrix diagonalization, which might become restrictive for large systems or high temperatures. 

In this Letter, we compute the electrical conductivity directly from the microscopic formulation of Ohm's law under the conditions found in Earth's core. This is achieved using the real-time formalism of TDDFT (RT-TDDFT)~\cite{PhysRevB.54.4484,castro2006octopus,ullrich2011time}.
By applying a weak external field, the electronic response, which determines optical properties and electronic transport properties, are extracted~\cite{bertsch2000real,baczewski2016x,andrade2018negative}. 
For certain regimes of electronic excitation and large systems, RT-TDDFT can be computationally more efficient than LR-TDDFT. As in LR-TDDFT, the response function computed using RT-TDDFT captures collective effects that are not captured in the standard approach using the KG formula.
As a central result, we provide an independent computational assessment, based on microscopic Ohm's law, of the yet unresolved controversy on the electrical conductivity of iron under the temperature and pressure conditions prevalent in Earth's core~\cite{dobson2016earth}. 
We compare our results with the experimental measurements by Ohta~\cite{ohta2016experimental} and Kon{\^o}pkov{\'a}~\textit{et al.}~\cite{konopkova2016direct}, as well as other computational results~\cite{gomi2013high,de2012electrical,pozzo2012thermal,pozzo2014thermal,PhysRevLett.121.096601,pourovskii2020electronic,PhysRevB.87.014110,he2022superionic}. 
Our analysis reveals better agreement with the electrical resistance measurements of Ohta \textit{et al.}~\cite{ohta2016experimental}. Furthermore, the temperature dependence of our results follows the trend observed in that experiment, improving significantly upon prior numerical investigations. 

\textit{Results.~---~} 
The microscopic formulation of Ohm's law describes how an external electric field $\boldsymbol{E}(\omega)$ gives rise to an induced electric current 
\begin{equation}
\label{eq:ohms.law}
\boldsymbol{J} (\omega) = \boldsymbol{\sigma}(\omega)\, \boldsymbol{E} (\omega)\,,
\end{equation}
where the constant of proportionality can be identified as the electrical conductivity $\boldsymbol{\sigma}(\omega)$. Note that Ohm's law is formulated in the frequency domain and that both the current and the electric field are vectors, while the conductivity is a tensor. Also note that we adopt Hartree atomic units, i.e., $\hbar = e = m_e = a_0 = 1$.

\begin{figure}[b]        
\centering       
\includegraphics[width=1.0\columnwidth]{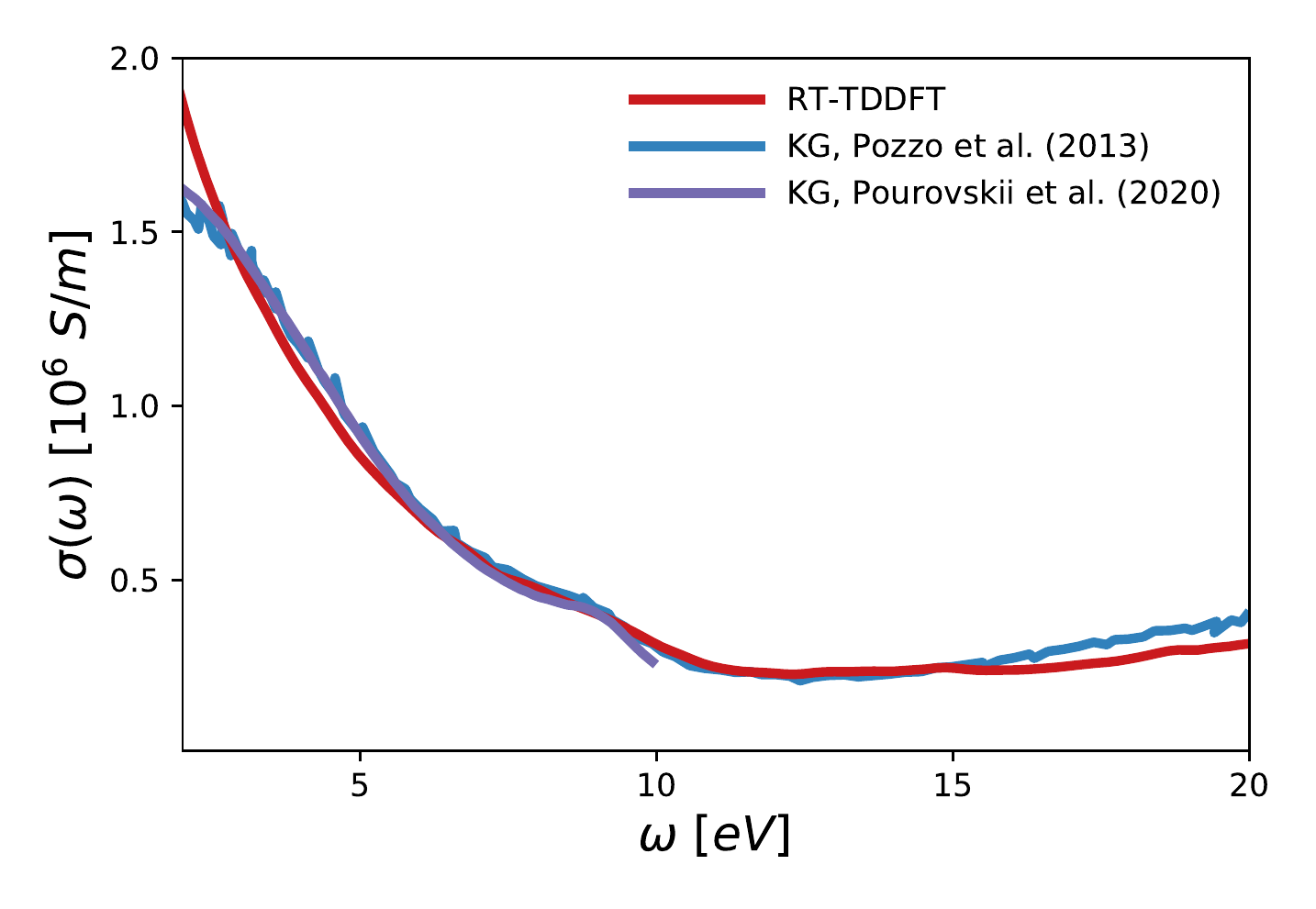}
\caption{Dynamical electrical conductivity under Earth core conditions ($T=6350$~K, $P=322$~GPa) from Ohm's law based on our RT-TDDFT calculations (red) compared with a previous work which uses the KG formula based on static DFT calculations~\cite{PhysRevB.87.014110} (blue) at a slightly higher pressure of 328 GPa and KG results~\cite{pourovskii2020electronic} (violet) at a lower temperature ($T=5802$~K) and slightly lower pressure ($P=310$~GPa).} 
\label{sigma_comp_pozzo} 
\end{figure} 

We compute the induced current on the atomistic level by using RT-TDDFT. By applying an electric field $\boldsymbol{E}(t) = -(1/c) (\partial \boldsymbol{A}/\partial t)$, where $\boldsymbol{A}$ is the impressed vector potential and $c$ is the speed of light, we obtain the induced time-dependent current density $\boldsymbol{j}(\br,t)=\Im [\sum_{i}^{N} \phi_{n,k}^{*}(\br,t) \nabla \phi_{n,k}(\br,t)] + n(\br,t) \boldsymbol{A}_S(\br,t)/c$. When integrated over the spatial coordinates,  it yields a time-dependent electric current $\boldsymbol{J}(t)$.  By taking the Fourier transform we obtain Ohm's law in the frequency domain as denoted in Eq.~(\ref{eq:ohms.law}).
The time-dependent current density is obtained by solving the time-dependent KS equations 
\begin{equation}
\label{eq:tdks}
\hat{H}_S \phi_{n,k}(\br,t)=i \frac{\partial}{\partial t} \phi_{n,k}(\br,t)
\end{equation}
for the KS orbitals $\phi_{n,k}(\br,t)$. Here, the effective Hamiltonian is given by    
\begin{equation}
\hat{H}_{S} = \frac{1}{2} \left[-i \nabla + \frac{1}{c} \boldsymbol{A}_S(\br,t)\right]^{2} + V_S(\br,t) \,
\end{equation}
where $V_S(\br,t)=V_{ext}(\br,t)+V_{H}(\br,t)+V_{XC}(\br,t)$ is the KS potential involving the sum of the external, the Hartree, and XC potentials, while the effective vector potential $\boldsymbol{A}_{S}(\br,t)=\boldsymbol{A}(\br,t) + \boldsymbol{A}_{XC}(\br,t)$ comprises the sum of the external vector potential and the XC contribution. The following RT-TDDFT results are obtained from an all-electron full-potential linearized augmented plane wave (FP-LAPW) method~\cite{singh2006planewaves} as implemented in the Elk~\cite{elk} and Exciting~\cite{gulans2014exciting,pela2021all} codes. For the sake of clarity and reproducibility, we provide a comprehensive description of all computational details and simulation parameters in the Supplemental Material~\cite{supplement}.      

We begin with computing the frequency-dependent response of the electrons in iron at a pressure of 322 GPa and a temperature of 6350 K as found in the Earth's core. To that end, we first prepare an appropriate initial electronic state. We follow the common procedure of generating uncorrelated atomic snapshots from Born-Oppenheimer molecular dynamics simulations based on static DFT at the given temperature and pressure. Here, our simulation cells contain 16 iron atoms. We assess the behavior of the radial distribution function and demonstrate liquid-like behavior under the given conditions (see Fig. 1 in the Supplemental Material~\cite{supplement}). Subsequently, we apply a step-like vector potential and solve Eq.~(\ref{eq:tdks}) for a duration of up to $t=$1000~a.u.. As commonly assumed in TDDFT, we invoke the adiabatic approximation which means that we neglect the temporal non-locality of the time-dependent KS potential and evaluate a ground-state XC functional on the density at time $t$. We follow this common procedure and employ the so-called adiabatic local density approximation~\cite{GrKo1985}. 
Based on the solutions $\phi_{n,k}(\br,t)$, we calculate the time-dependent current density $\boldsymbol{j}(\br,t)$ which we integrate over the spatial coordinates to obtain the electric current $\boldsymbol{J}(t)$. The frequency-dependent, i.e., dynamical electrical conductivity is then extracted from the electric current based on Ohm's law as given in Eq.~(\ref{eq:ohms.law}). Care has to be taken in the choice of parameters for the external vector potential and for the Fourier transform of the macroscopic current from the time to the frequency domain. These details along with the choice of computational and methodological parameters are also included in the Supplemental Material~\cite{supplement}. We then compute the dynamical electrical conductivity under the Earth core conditions. Fig.~\ref{sigma_comp_pozzo} illustrates the result of our RT-TDDFT calculations (red curve) with an energy resolution of 0.17~eV which is proportional to the inverse of the total propagation time. The calculations converge quickly given a sufficient set of KS orbitals, even for a modest size of the supercell.  

We compare our calculations with prior results obtained from using the KG formula based on static DFT~\cite{PhysRevB.87.014110,pourovskii2020electronic}. In this particular case, all methods yield similar results except for a discrepancy in the $\omega \rightarrow 0$ limit which corresponds to the DC conductivity. Here, the KG results are more susceptible to finite-size effects and are very sensitive to the location and density of the KS eigenvalues. The results by Pozzo \textit{et al.} are for a system consisting of 157 atoms and those by Pourovskii \textit{et al.} are for 250 atoms. 

\begin{figure}[th]       
\centering       
\includegraphics[width=1.0\columnwidth]{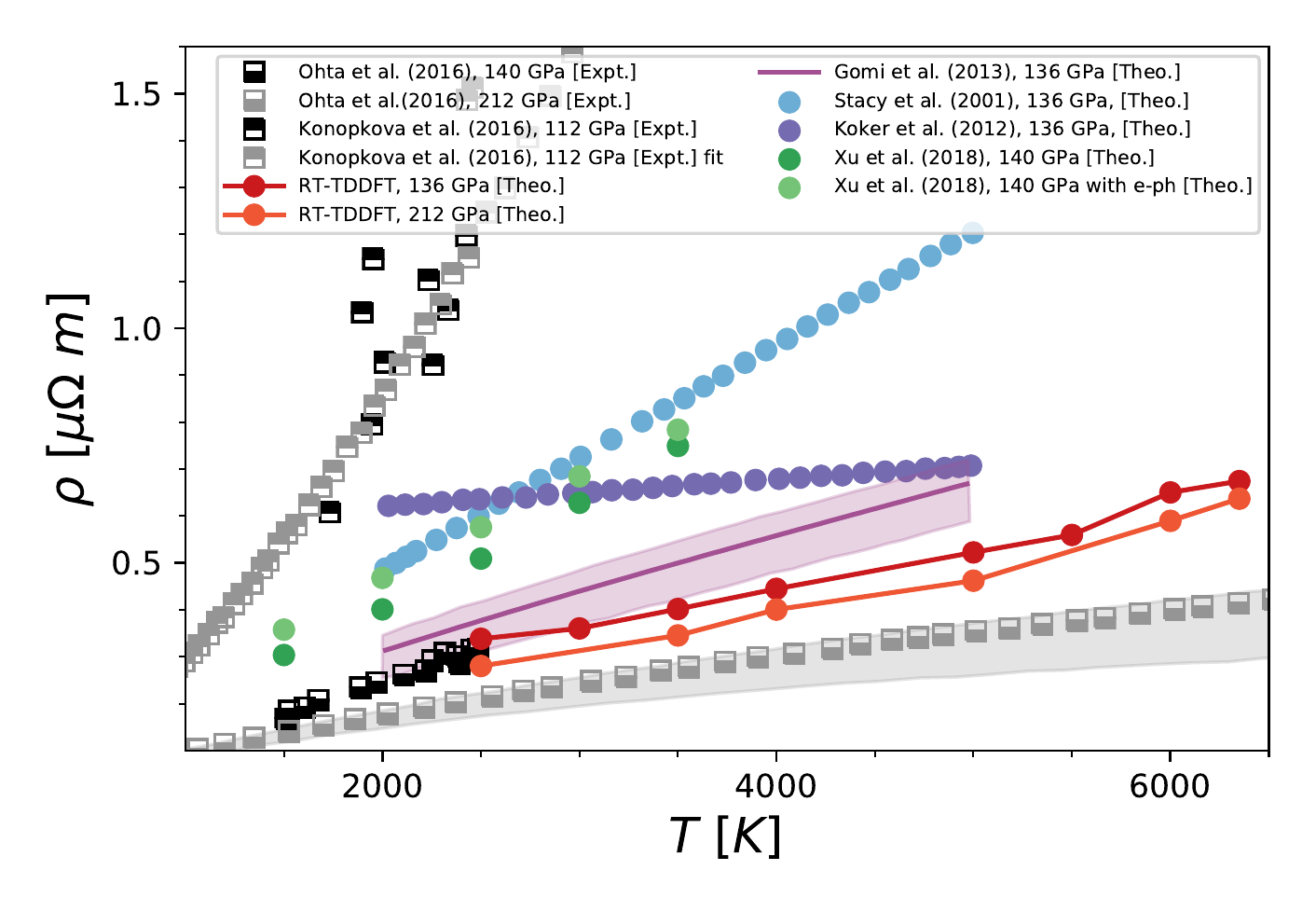} 
\caption{Electrical resistivity and its temperature dependence at 136($\pm5$)~GPa and 212($\pm6$)~GPa.
Results of diamond anvil cell measurements were reported by Ohta \textit{et al.}~\cite{ohta2016experimental} (lower filled black and grey squares) and by Kon{\^o}pkov{\'a} \textit{et al.}~\cite{konopkova2016direct} (upper filled black and grey squares). The electrical resistivity predicted by Ohm's law based on our RT-TDDFT calculations (red circles) are compared with first-principles calculations by Xu \textit{et al.}~\cite{PhysRevLett.121.096601} (green circles), interpolated results of Stacey \textit{et al.}~\cite{stacey2001electrical} (blue curve), a modified Bloch-Gr\"{u}neisen model of Koker \textit{et al.}~\cite{de2012electrical} (violet curve), and a Bloch-Gr\"{u}neisen model including resistivity saturation by Gomi \textit{et al.}~\cite{gomi2013high} (purple curve).}
\label{rho_comp_keeler}   
\end{figure} 
Next, we come to the central result presented in Fig.~\ref{rho_comp_keeler} where we compare the predictions of our RT-TDDFT calculations with the discordant experimental measurements reported by Ohta \textit{et al.}~\cite{ohta2016experimental} and Kon{\^o}pkov{\'a} \textit{et al.}~\cite{konopkova2016direct}. 
In order to compare our calculations with the reported experiments, we extrapolate the dynamic conductivity to $\omega \rightarrow 0$ limit.

Shown in Fig.~\ref{rho_comp_keeler} is the behavior of the electrical resistivity $\rho$ (the inverse of the conductivity) as a function of the temperature at fixed, high pressures. The experimental DAC measurements reported by Ohta \textit{et al.}~\cite{ohta2016experimental} (lower filled black and grey squares) are contrasted with those by Kon{\^o}pkov{\'a} \textit{et al.}~\cite{konopkova2016direct} (upper filled black squares). Note that the data by Kon{\^o}pkov{\'a} \textit{et al.} are based on their thermal conductivity measurements which we have converted into an electrical resistivity using the Wiedemann-Franz law~\cite{Wiedemann.Franz.law} with a Lorenz number 2.44$\times 10^{-8}$~W$\Omega$K$^{-2}$. In addition, their fit (upper filled grey squares) to the experimental data is also shown.
The proportionality $\rho \propto T$ at these conditions is observed in other results too that are described by a Bloch-Gr\"{u}neisen model based on the Debye temperature~\cite{jaccard2002superconductivity}. The net effect of pressure is to  decrease the resistivity as the smaller amplitude of the ionic vibrations are responsible for the increase in the mean free path of the electrons. The striking feature of this plot is that the electrical resistivity predicted by Ohm's law based on our RT-TDDFT calculations agrees well with the measurements of Ohta \textit{et al.}, particularly with the data points at a pressure of 140 GPa and a temperature of 2500 K (red circles).
Contrarily, other previously reported calculations based on other methods did not favor any of the two experiments with the exception of the Bloch-Gr\"{u}neisen model including resistivity saturation by Gomi \textit{et al.}~\cite{gomi2013high} (purple curve). Note that the Bloch–Grüneisen model describes the effect of temperature on resistivity for a given volume and is not a \textit{first principles} result. Other prior works include the interpolated results of Stacey \textit{et al.}~\cite{stacey2001electrical} (blue curve), the Bloch-Gr\"{u}neisen model of Koker \textit{et al.}~\cite{de2012electrical} (violet curve), and the first-principles calculations by Xu \textit{et al.}~\cite{PhysRevLett.121.096601} including the electron-phonon contribution (green circles). Furthermore, the contribution of EES in HCP iron to the resistivity under Earth core conditions is well assessed by Pourovskii \textit{et al.}~\cite{pourovskii2017electron} in terms of DMFT leading to the behavior $\rho_{EES} \propto T^{2}$. The effects of EES are negligible at the data points of 2500$-$3000 K in this work. 

\begin{figure}[th]          
\centering       
\includegraphics[width=1.0\columnwidth]{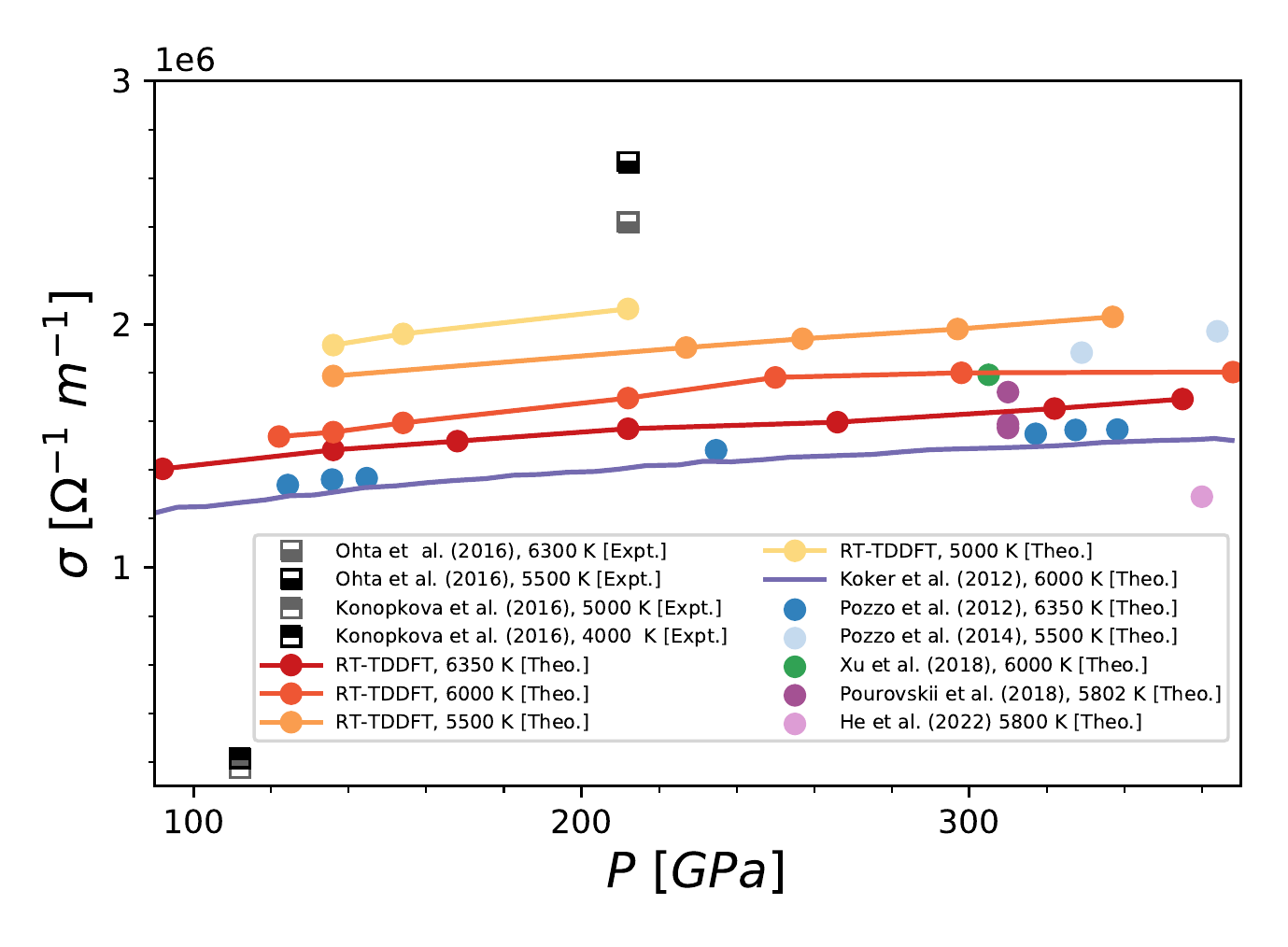} 
\caption{Electrical conductivity and its pressure dependence at fixed temperature. Results of the experimental diamond anvil cell measurements reported by Ohta \textit{et al.}~\cite{ohta2016experimental} (lower filled black and grey squares) are contrasted with those by Kon{\^o}pkov{\'a} \textit{et al.}~\cite{konopkova2016direct} (upper filled grey squares). Our predictions of the electrical conductivity (red circles) are compared with previously reported calculations, such as the Bloch-Gr\"{u}neisen model of Koker \textit{et al.}~\cite{de2012electrical} (violet curve), the KG formula by Pozzo \textit{et al.}~\cite{pozzo2012thermal,pozzo2014thermal} (blue circles), first-principles calculations including electron-phonon contributions by Xu \textit{et al.}~\cite{PhysRevLett.121.096601} (green circle), and first-principles calculations including electron-electron and electron-lattice scattering by Pourovskii \textit{et al.}~\cite{pourovskii2020electronic} (purple circles) and He \textit{et al.}~\cite{he2022superionic} (pink circles).}   
\label{sigma_vs_pressure_with_temp_innercore} 
\end{figure} 
Finally in Fig.~\ref{sigma_vs_pressure_with_temp_innercore}, we consider the electrical DC conductivity as a function of the pressure at various fixed temperatures up to the Earth core conditions. While not as striking as in Fig.~\ref{rho_comp_keeler}, our RT-TDDFT predictions (red, orange, light orange, and yellow circles) are closer to the experimental results by Ohta \textit{et al.} (lower filled black and grey squares) then to those by Kon{\^o}pkov{\'a} \textit{et al.} (upper filled grey squares). Again, we used the Wiedemann-Franz law~\cite{Wiedemann.Franz.law} to extract the DC conductivity from the experimental data by Kon{\^o}pkov{\'a} \textit{et al.}. Furthermore, when compared with prior simulation results, our RT-TDDFT results are in better agreement with those by Ohta \textit{et al.} throughout. 
We also compare with results obtained from the Bloch-Gr\"{u}neisen model of Koker \textit{et al.}~\cite{de2012electrical} (violet curve), the KG formula by 
Pozzo \textit{et al.}~\cite{pozzo2012thermal,PhysRevB.87.014110,pozzo2014thermal} (blue circles), density functional perturbation theory combined with the Korringa-Kohn-Rostoker method~\cite{korringa1947calculation,kohn1954solution} that include electron-phonon contributions by Xu \textit{et al.}~\cite{PhysRevLett.121.096601} (green circle), and the dynamical mean field calculations which also capture EES in the BCC and HCP phases of iron by Pourovskii \textit{et al.}~\cite{pourovskii2020electronic} (purple circles) and similarly for the HCP phase only by He \textit{et al.}~\cite{he2022superionic} (pink circle). Overall, the change in the conductivity with pressure is predicted to be relatively small by all models and theories. 

We conclude this investigation of electronic transport properties by providing a concise assessment of the methods discussed here. We point out that calculations using the KG formula do not include an interaction kernel and, thus, do not take into account collective effects like plasmons. LR-TDDFT is an extension of the KG formula in terms of an interaction kernel. Both the KG formula and LR-TDDFT are often limited to the head of the density response matrix and, therefore, neglect so-called local field effects originating from the off-diagonals. In RT-TDDFT, however, we make no such assumption. Both the complete electronic response and the interaction kernel in terms of Hartree and XC contributions are considered~\cite{baczewski2021predictions}.

\textit{Conclusions.~---~} 
In this letter, we have reported results on the electrical conductivity of iron under the conditions of Earth's core from the microscopic formulation of Ohm's law. The yet unresolved disagreement between the experimental results of Ohta \textit{et al.}~\cite{ohta2016experimental} and Kon{\^o}pkov{\'a} \textit{et al.}~\cite{konopkova2016direct} demonstrates the level of uncertainty in our understanding of the transport properties and thus in the dynamo action and related quantities of Earth's core. While previous simulation efforts remained inconclusive, we obtain a much better agreement with the experimental results of Ohta \textit{et al.}. Thereby, we provide an independent assessment of previously reported simulation results. We demonstrate the utility of our method, which is based on the real-time formalism of time-dependent DFT, for computing transport properties in materials under extreme conditions. It provides a viable alternative to current state-of-the-art methods, such as the evaluation of the KG formula on DFT data. We expect our method to become a widely used device for the interpretation of upcoming free-electron laser scattering experiments at facilities like LCLS~\cite{Fletcher2015}, the European X-FEL~\cite{Tschentscher_2017}, and FLASH~\cite{zastrau_resolving_2014}.
While in this work, the perturbing vector potential was chosen in the linear regime, our method is also valid in the non-linear regime. This will enable studying the response of materials under extreme conditions accessible through recent advances in free-electron lasers~\cite{zastrau2021high,cerantola2021new}. 

\textit{Acknowledgements.~---~} 
The authors acknowledge Thomas R. Mattsson for his motivation to compare the predictions of TDDFT transport properties with the experimental work on iron under the conditions of Earth's core. This work was partially supported by the Center of Advanced Systems Understanding (CASUS) which is financed by Germany’s Federal Ministry of Education and Research (BMBF) and by the Saxon state government out of the State budget approved by the Saxon State Parliament.
ML was supported by the German Federal Ministry of Education and Research (BMBF, No. 01/S18026A-F) by funding the competence center for Big Data and AI “ScaDS.AI Dresden/Leipzig.”  Computations were performed on a Bull Cluster at the Center for Information Services and High Performance Computing (ZIH) at Technische Universit\"at Dresden, on the cluster Hemera at Helmholtz-Zentrum Dresden-Rossendorf (HZDR). 
Sandia National Laboratories is a multimission laboratory managed and operated by National Technology and Engineering Solutions of Sandia, LLC, a wholly-owned subsidiary of Honeywell International Inc., for the U.S. Department of Energy’s National Nuclear Security Administration under contract DE-NA0003525.


\bibliography{bibliography}

\end{document}